\input{aipcheck}

\documentclass[
    ,final            
  ]
  {aipproc}

\layoutstyle{6x9}

\def\beqn{\begin{eqnarray}}
\def\eeqn{\end{eqnarray}}
\def\barr{\begin{array}}
\def\earr{\end{array}}
\def\btab{\begin{tabular}}
\def\etab{\end{tabular}}
\def\bite{\begin{itemize}}
\def\eite{\end{itemize}}
\def\bcen{\begin{center}}
\def\ecen{\end{center}}

\def\eq{\begin{equation}}
\def\ee{\end{equation}}
\def\nn{\nonumber}

\def\q2dagger{q_2\hspace{-0.35cm}/\;}

\begin{document}
\title{Dispersion corrections to parity violating electron scattering}
\author{M. Gorchtein}{
  address={Nuclear Theory Center, Indiana University, Bloomington, IN 47408, USA}
}
\author{C. J. Horowitz}{
  address={Nuclear Theory Center, Indiana University, Bloomington, IN 47408, USA}
}
\author{M. J. Ramsey-Musolf}{
  address={University of Wisconsin-Madison, Madison, WI 53706, USA}
 }
\begin{abstract}
We consider the dispersion correction to elastic parity violating 
electron-proton scattering due to $\gamma Z$ exchange. In a recent publication, 
this correction was reported to be substantially larger than the previous 
estimates. In this paper, we study the dispersion correction in greater detail. 
We confirm the size of the disperion correction to be $\sim$6\% for
the QWEAK experiment designed to measure the proton weak charge. 
We enumerate parameters that have to be
constrained to better than relative 30\% in order to keep the
theoretical uncertainty for QWEAK under control.
\end{abstract}
\maketitle
Precision tests of the Standard Model at low energies provide an
important framework for New Physics searches and for setting stringent
constraints on the parameters of possible extensions of the Stadard
Model (SM). Such tests involve high precision measurements of parameters 
that are suppressed or precisely vanish in SM. 
An important example of such parameter is the weak charge 
of the proton, $Q_W^p=1-4\sin^2\theta_W$.  With the value of the weak mixing 
angle at low momentum transfers $\sin^2\theta_W(0)=0.23807\pm0.00017$ 
\cite{erler}, the SM predicts the proton weak charge of order $\approx0.05$. 
 A precise 4\% (combined 2\% experimental and 2\% theoretical uncertainties)
determination of the weak charge of the proton is the aim of the QWEAK experiment 
at Jefferson Lab \cite{qweak}.  In this experiment, parity violating
asymmetry in polarized electron scattering $A_{PV}=(\sigma_R-\sigma_L)/
(\sigma_R+\sigma_L)$ will be measured, $\sigma_R(\sigma_L)$
standing for the differential cross section in elastic electron-proton
scattering with the incidental electron beam spin parallel
(antiparallel) to its direction, respectively. 
To leading order in the momentum transfer $t$ and in Fermi constant, 
the parity violating asymmetry arises from the interference of parity
conserving and parity violating amplitudes at tree level. 
Including radiative corrections of order $\alpha$, one has
\beqn
A^{PV}=\frac{G_Ft}{4\pi\alpha\sqrt{2}}
Q_W^p[1+{\rm Re}\delta_{RC}+{\rm Re}\delta_{\gamma Z}(\nu)]+{\cal{O}}(t^2)\,,
\eeqn
\indent
Above, we denote all the radiative 
corrections except the $\gamma Z$ direct and crossed boxes 
by $\delta_{RC}$, and we explicitly indicate that the dispersion
correction $\delta_{\gamma Z}$ is function of energy $\nu$. 
The corrections $\delta_{RC}$ were considered in various works, 
to mention the most important references \cite{erler,sirlin}, and the 
combined theoretical uncertainty associated with these 
was shown not to exceed 2.2\%. 
The dispersion correction $\delta_{\gamma Z}$ is the only one that can obtain a sizeable
contribution from hadronic structure. 
In \cite{gammaZ}, the dispersion correction was represented as the sum 
$\delta_{\gamma {Z}}=\delta_{\gamma {Z_A}}+\delta_{\gamma {Z_V}}$, and two forward sum rules 
were established,
\beqn
{\rm Re}\delta_{\gamma {Z_A}}(\nu)&=&\frac{2\nu}{\pi}\int_{\nu_\pi}^\infty
\frac{d\nu'}{\nu'^2-\nu^2}{\rm Im}\delta_{\gamma{Z_A}}(\nu')\,,
\label{dr:Z_A}\\
{\rm Re}\delta_{\gamma {Z_V}}(\nu)&=&\frac{2}{\pi}\int_{\nu_\pi}^\infty
\frac{\nu'd\nu'}{\nu'^2-\nu^2}{\rm Im}\delta_{\gamma{Z_V}}(\nu')\,,
\label{dr:Z_V}
\eeqn
\indent
The respective imaginary parts can be expressed as integrals over 
the PV DIS structure functions $\tilde F_i(x,Q^2)$,
\beqn
{\rm Im}\delta_{\gamma {Z_A}}(\nu)&=&\frac{\alpha}{2Q_W^p}g_A^e
\int_{W^2_\pi}^s\frac{dW^2}{(s-M^2)^2}
\int_0^{Q^2_{max}}\frac{dQ^2}{1+\frac{Q^2}{M_Z^2}}
\left[\tilde{F}_1
+ \left(\frac{2Pk_1Pk}{PqQ^2}-\frac{P^2}{2Pq}\right)\tilde{F}_2\right]\nn\\
{\rm Im}\delta_{\gamma {Z_V}}(\nu)&=&-\frac{\alpha}{2Q_W^p}g_V^e
\int_{W^2_\pi}^s\frac{dW^2}{(s-M^2)^2}
\int_0^{Q^2_{max}}\frac{dQ^2}{1+\frac{Q^2}{M_Z^2}}
\frac{(P,k+k_1)}{2(Pq)}\tilde{F}_3
\label{imdeltagz}
\eeqn
with $W^2_\pi=(M+m_\pi)^2$ the pion production threshold, and 
$Q^2_{max}=\frac{(s-M^2)(s-W^2)}{s}$.
The sum rules of Eqs. (\ref{dr:Z_A},\ref{dr:Z_V}) are new results. In
the past, only their values at $\nu=0$ were calculated in the
framework of atomic PV. Due to the explicit factor of $\nu$, the
correction $\delta_{\gamma Z_A}$ is exactly zero for zero energy, in
accordance with \cite{sirlin}, and has been neglected in the
literature even for non-zero energies. The dispersion
relation calculation of \cite{gammaZ} lead to $\delta_{\gamma Z_A}\sim6\%$
that should be compared with zero in the analysis of the QWEAK
experiment. 
The correction 
$\delta_{\gamma Z_V}$ obtains its value due to hard kinematics inside
the loop and is largely energy-independent \cite{sirlin,mjrm}. In the
rest of this article, we concentrate on $\delta_{\gamma Z_A}$ only.
The sum rule of Eq. (\ref{dr:Z_A}) 
itself is model-independent; however, in absence of any detailed 
PVDIS data, the input in this sum rule will depend on a model. 
We proceed by modeling the electromagnetic data first.
\section{Modeling Real and virtual photoabsorption data}
\begin{figure}
\includegraphics[width=4in]{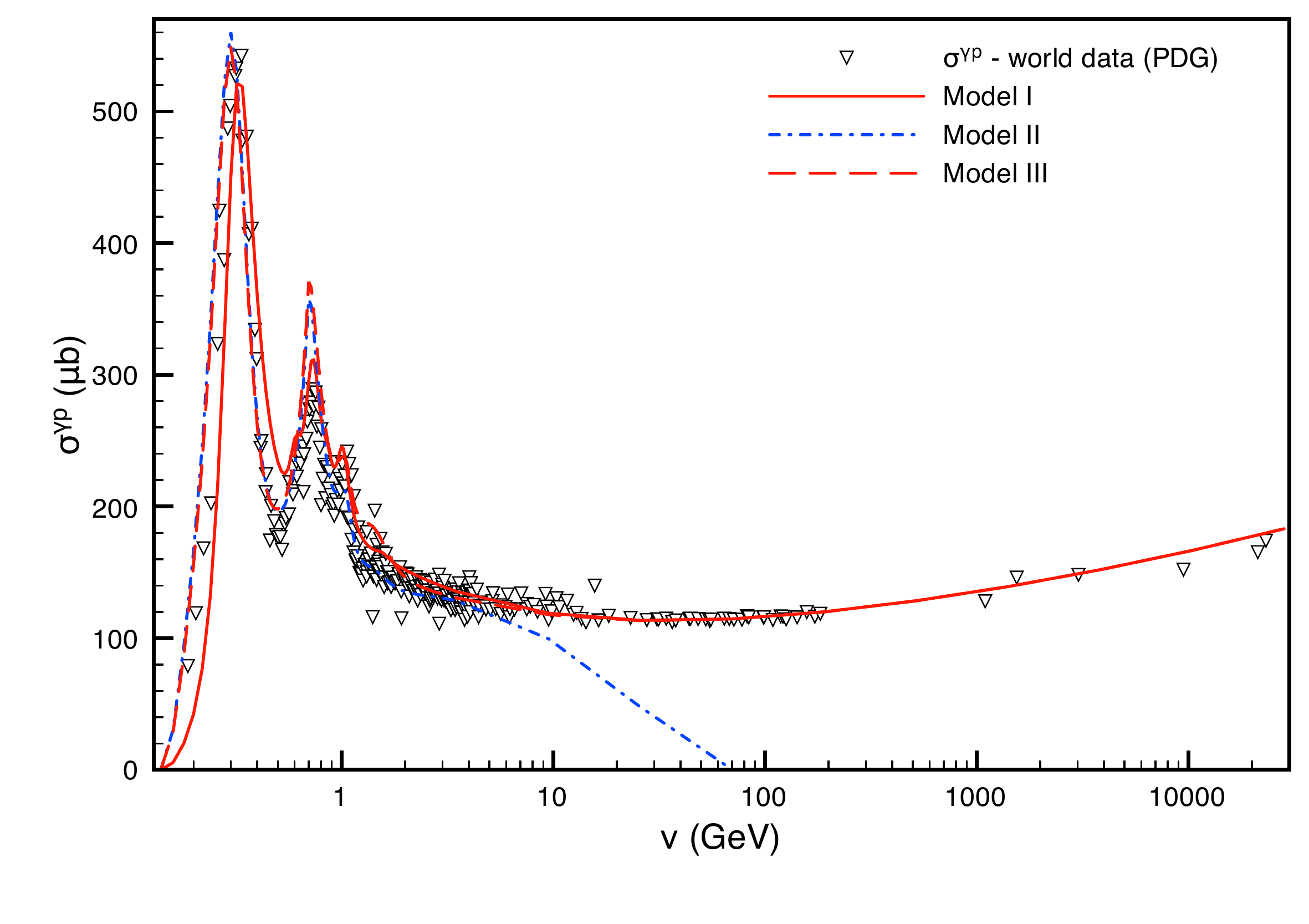}
\caption{World data on total photoabsorption 
\cite{bloom69,armstrong72,caldwell73,caldwell78,chekanov02,vereshkov03}
(see \cite{pdg} for the complete list) compared to
  the three models described in the text.} 
\label{fig:sigmatot}
\end{figure}

\begin{figure}
\includegraphics[width=5in]{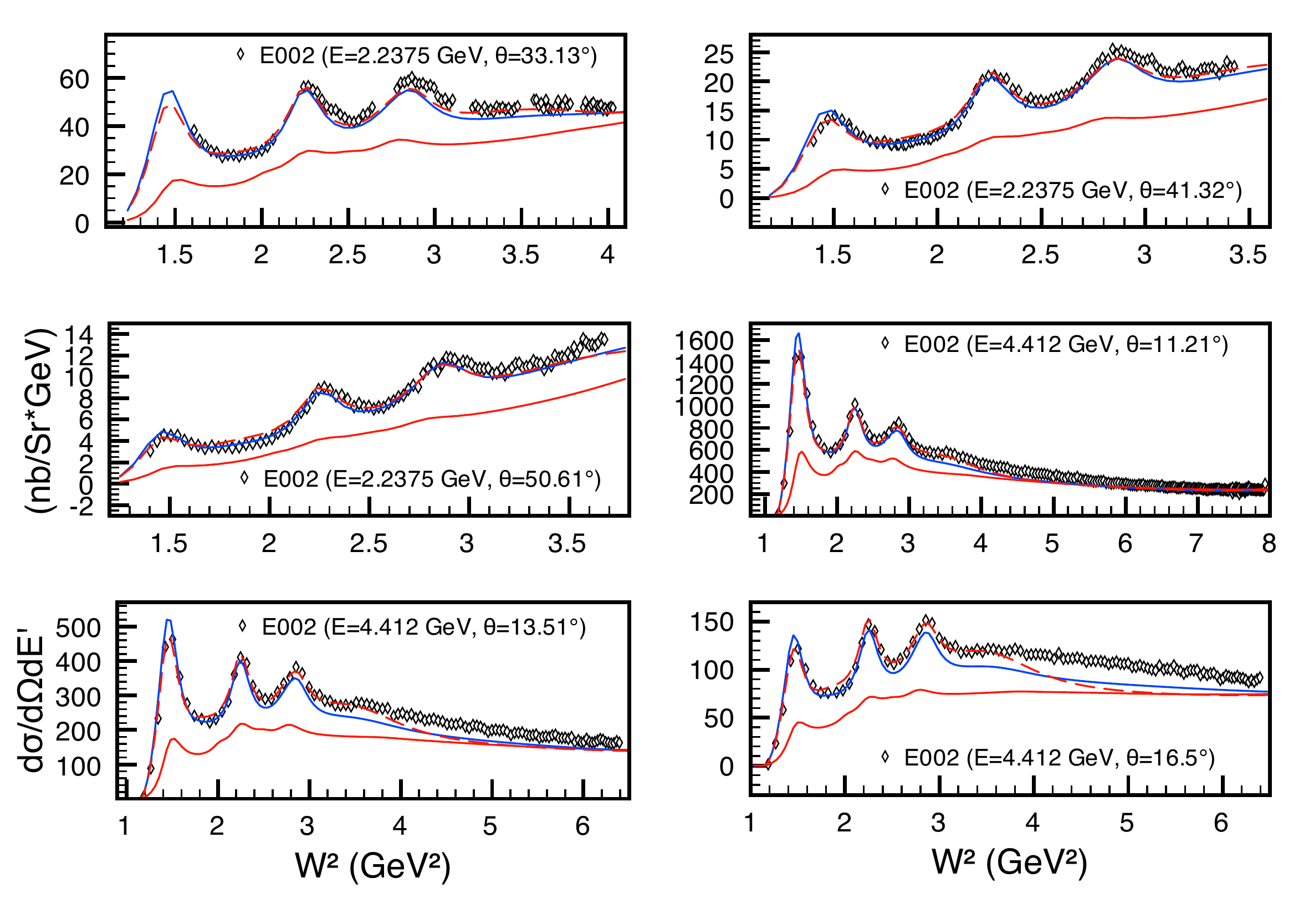}
\caption{Differential cross section data in the resonance region from
  JLab E002 \cite{E002} are shown in comparison with the  three models.} 
\label{fig:resdata1}
\end{figure}

We will use the following three models to model the electromagnetic
DIS structure functions $F_i$:
\begin{itemize}
\item
{\bf Model I}: 
The model used in \cite{gammaZ} utilized the resonance parameters
obtained in \cite{bianchi} and the non-resonant Regge contribution
from \cite{cvetic} that was fitted to the real photon data at
high energies. Ref. \cite{cvetic} also provides the $Q^2$-dependence
of the high-energy part, so no additional modelling was necessary
here. For the estimates of \cite{gammaZ}, a simple dipole model with
the dipole mass $\Lambda\approx1$ GeV for all the transition resonance 
form factors was employed. 
\item
{\bf Model II}: 
Another form of resonance and background contributions and transition
form factors was used in \cite{bosted} to fit virtual photon data. 
\item
{\bf Model III}: 
As Fig. \ref{fig:sigmatot} clearly demonstrates, the background of
\cite{bosted} cannot be used too far beyond the resonance region. 
Therefore, we opt for Model III that uses 
the background from Model I, and adopt the resonance contribution from Model II.
\end{itemize}
In Figs. \ref{fig:sigmatot} and \ref{fig:resdata1}, we display the comparison
of the three models with the data for the differential cross section for
inclusive electroproduction in the resonance region. 
The Model I \cite{gammaZ} is shown by solid red lines, Model II 
by the solid blue line, and Model III is represented by a long-dash red line. 
From Fig. \ref{fig:resdata1} it becomes clear that
Model I largely underestimates the JLab data in the whole resonance
region, apart from the real photon point Fig. \ref{fig:sigmatot}. 
Model III also provides a very good description of the DIS data for
$F_2(x,Q^2)$ at low and moderate values of $Q^2$ \cite{cvetic}. 
\section{Isospin structure}
Once the parameters are fixed, we can employ simple isospin
considerations for each contribution. Note that because breaking down
the experimental data into the sum of resonances and background is a
model-dependent procedure, also the isospin rotation is only defined
within a given model. For the $N\to N^*(I=1/2)$ transition, the
isospin decomposition resembles that for the elastic form factors, 
$\langle N^*|J^\mu_{NC,V}|p\rangle=(1-4s^2\theta_W) \langle N^*|J^\mu_{em}|p\rangle-\langle N^*|J^\mu_{em}|n\rangle $
It is then straightforward to relate the contribution of a resonance
$R$ with isospin $1/2$ to the interference $\gamma Z$
"cross section", to its contribution to the electromagnetic cross
section by introducing the isospin scaling factors
\beqn
\xi_{Z/\gamma}^R\equiv\frac{\sigma_{T,R}^{\gamma Z,p}}
{\sigma_{T,R}^{\gamma\gamma p}}&=&(1-4s^2\theta_W)
-\frac{ A_{R,1/2}^p A_{R,1/2}^{n*}+A_{R,3/2}^p A_{R,3/2}^{n*} }
{|A_{R,1/2}^p|^2+|A_{R,3/2}^p|^2}
\eeqn
\indent
Above, $ A_{R,1/2(3/2)}^{p(n)}$ are the transition helicity amplitudes
for exciting the resonance $R$ on the proton (neutron), respectively,
whereas $\sigma_{T,R}$ denotes the contribution of the resonance $R$ 
to the transverse virtual photon cross section. 
The scaling factors $\xi_{Z/\gamma}^R$ depend both on the
relative size and relative phase of the transition amplitude on the
proton and on the neutron. To obtain both, we use the results of the 
constituent quark model of Ref. \cite{koniuk} 
for the transition helicity amplitudes. For isovector transitions $N\to\Delta(\Delta^*)$, 
the scaling factor in the SM is $\xi_{Z/\gamma}^\Delta=2(1-2\sin^2\theta_W)$.
Vector Meson Dominance Model (VDM) capitalizes on the fact that the
photon ($Z$-boson) has the same quantum numbers as vector mesons and
can be represented as a superposition of a few vector mesons,
$|\gamma\rangle=\sum_{V=\rho,\omega,\phi,\dots}|V\rangle$.
In the naive VDM, these three channels cannot mix among each other,
and one obtains a prediction for the ratios of the cross sections of
$\rho,\,\omega,\,\phi$ production, 
$\sigma^{\gamma^*p\to\rho p}\;:\;
\sigma^{\gamma^*p\to\omega p}\;:\;
\sigma^{\gamma^*p\to\phi p}=
1\;:\;(q^{I=0}/q^{I=1})^2\;:\;
(q^{s}/q^{I=1})^2=1\;:\;1/9\;:\;2/9$,
where we defined $q^{I=0}=1/(3\sqrt{2})$,
$q^{I=1}=1/\sqrt{2}$ and $q^s=-1/3$. 
The above predictions were confronted to the experimental data at high
energies and for $Q^2$ that ranged from zero to several GeV$^2$
\cite{rho_omega_phi} and showed a very good agreement for the
$\omega/\rho$ ratio, while for the $\phi/\rho$ ratio the agreement was
not as good. 
Accomodating these considerations here, we obtain the following ratio
of the high energy non-resonant (NR) contributions to $\gamma^*p\to Zp$
and $\gamma^*p\to\gamma^*p$ cross sections:
\beqn
\xi_{Z/\gamma}^{NR}
&\approx&
\frac{g_V^{I=1}q^{I=1}+g_V^{I=0}q^{I=0}+g_V^{s}q^{s}}{(q^{I=1})^2+(q^{I=0})^2+(q^{s})^2}
=2(1-4\sin^2\theta_W)\label{eq:xiNR}
\eeqn
where we make use of the SM 
definition $g_V^{I=0}=-4\sin^2\theta_W/(3\sqrt{2})$,
$q^{I=1}=(2-4\sin^2\theta_W)/\sqrt{2}$ 
and $q^s=-1+4\sin^2\theta_W/3$. 
We summarize these results inTable \ref{tab}.
\begin{table}[ht]
\vspace{1cm}
   \begin{tabular}{|l|l|l|l|l|l|l|l|}
\hline
$P_{33}(1232)$ & $S_{11}(1535)$ & $D_{13}(1520)$ & $S_{11}(1665)$ &
$F_{15}(1680)$ & $P _{11}(1440)$& $F _{37}(1950)$ & NR\\
\hline
1.075 & 0.885 & 0.938 & 0.473 & 0.35 & 0.745 & 1.075 & 1.075\\
\hline
   \end{tabular}
\caption{Isospin scaling factors $\xi_{Z/\gamma}$ for resonances and
  non-resonant (NR) background.}
\label{tab}
\end{table}
\section{Results for ${\rm Re}\,\delta_{\gamma Z_A}$}
\begin{figure}
\vspace{-1cm}
  \includegraphics[height=.3\textheight]{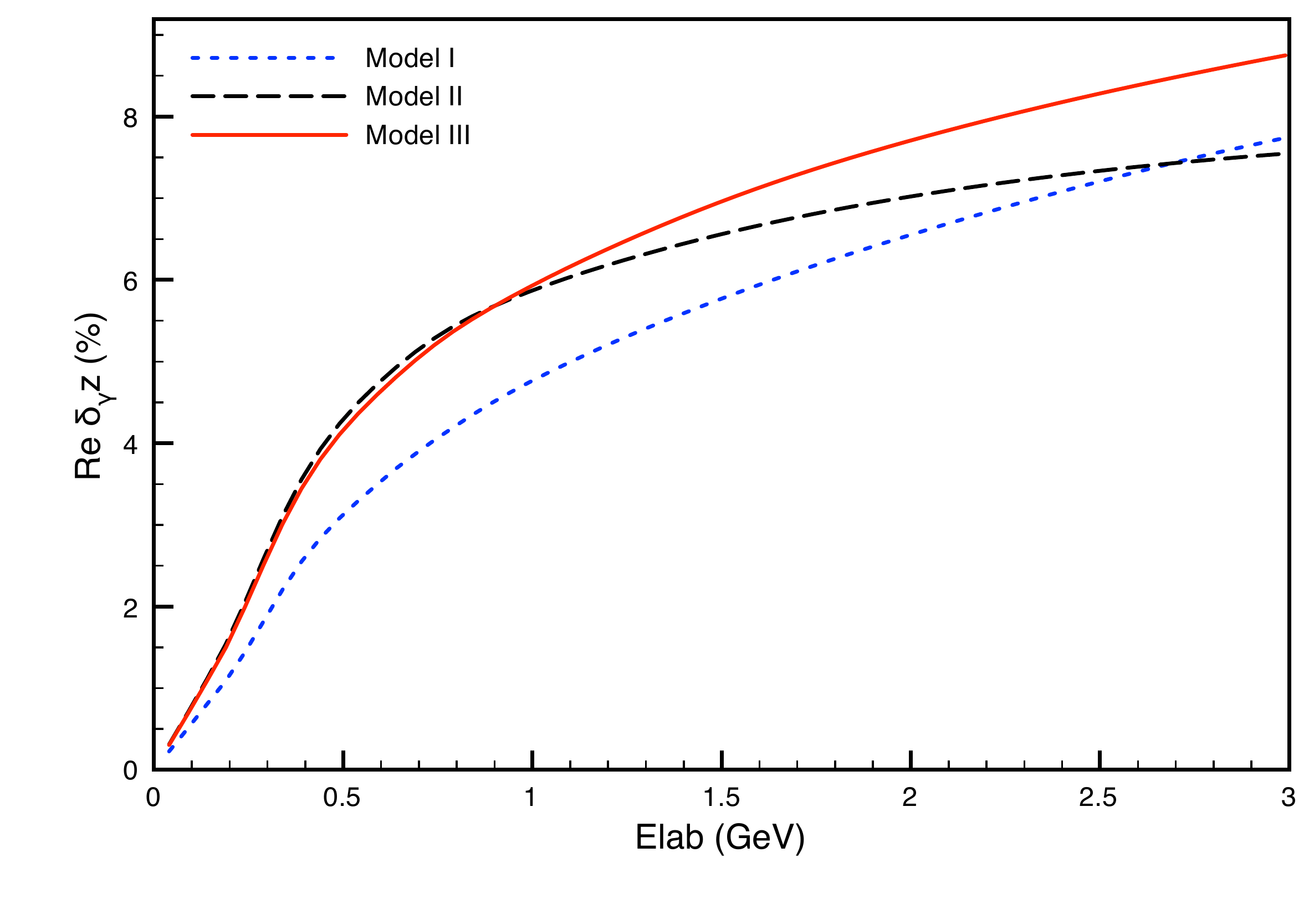}
\vspace{-1cm}
  \caption{Results for the dispersion correction as calculated in the
    Models I, II and III. }
\label{fig:deltagZ}
\end{figure}
Fig. \ref{fig:deltagZ} shows the energy dependence of ${\rm
  Re}\,\delta_{\gamma Z_A}$, calculated in three different models. The
individual contributions at QWEAK energy are shown in Table
\ref{tab2}. 
\begin{table}
   \begin{tabular}{|l|l|l|l|l|l|l|l|}
\hline
$P_{33}(1232)$ & $S_{11}(1535)$ & $D_{13}(1520)$ & $F_{15}(1680)$ & $S_{11}(1665)$ & $P _{11}(1440)$& $F _{37}(1950)$ & NR\\
\hline
1.27 \% & 0.44 \% & 0.21 \% & 0.06 \% & 0.05 \% & 0.10 \% & 0.50 \%  &  3.61 \%\\
\hline
   \end{tabular}
\caption{Individual resonances and background (NR) contributions to the
  dispersion correction Re$\delta_{\gamma Z_A}$ for QWEAK energy $E_{lab}=1.165$
  GeV. Results of Model III are quoted.}
\label{tab2}
\end{table}
We next discuss the size and origin of the uncertainty in calculating
Re$\,\delta_{\gamma Z_A}$. It comes about from i)
modeling the electromagnetic data, and ii) from the respective isospin
structure. To estimate the former, we average over the three models
discussed above and displayed in Fig. \ref{fig:deltagZ}, leading to 
Re$\delta_{\gamma Z_A}=(5.85\pm0.45)$\%, with most of the error bar
due to the Model I. Because the latter underestimates the experimental
data throughout the resonance region for non-zero $Q^2$, it is clear
that this estimate represents an upper bound for the model
dependence. In an upcoming work, we plan to directly fit the
parameters of Model III to the data that will allow to significantly 
reduce the model-dependent error, given the quality of the
experimental data shown in Figs. \ref{fig:sigmatot},
\ref{fig:resdata1}. 
The isospin structure represents a more significant source of
uncertainty. We will next indicate, to what precision the isospin
scaling factors of Table \ref{tab} should be constrained to keep the
uncertainty of Re$\,\delta_{\gamma Z_A}$ under control. 
From Table \ref{tab2}, it is seen that contributions of
$D_{13}(1520)$, $F_{15}(1680)$, $S_{11}(1665)$ and  $P _{11}(1440)$ resonances have
very little impact on the size of the dispersion correction. 
We can therefore afford to assign a conservative uncertainty of 50\% to each.
$P_{33}(1232)$ has been studied in many phenomenological
models, and we believe that it is realistic to constrain the scaling factor
$\xi_{Z/\gamma}^{P_{33}(1232)}$ to better than 20\% that would translate
into 0.25\% for the dispersion correction. For
$S_{11}(1535)$, the isospin structure might be more complex,
since it couples strongly to $\pi\pi N$ states along with $\pi N$, but
the absolute size of the correction quoted in Table \ref{tab2} allows
for a larger relative uncertainty here. Constraining
$\xi_{Z/\gamma}^{S_{11}(1535)}$ to relative 30\%  would lead to 0.13\% for $\delta_{\gamma Z}$. 
Finally, we turn to the uncertainties associated with
the higher energy contributions. It has to be mentioned that the contribution
associated in \cite{bosted} with a resonance around $W=1950\,$MeV cannot
be ultimately identified with the $F_{37}(1950)$ state listed in PDG
\cite{pdg}. This contribution may represent a missing strength of the background 
at $W\geq1900\,$MeV that can be seen in the two lower panels of 
Fig. \ref{fig:resdata1}, rather than a real resonance. Constraining
this contribution to relative 50\% means 0.25\%  for
$\delta_{\gamma Z}$. For the background, we assign the
100\% uncertainty to the strange quarks contribution to the scaling
factor in Eq. (\ref{eq:xiNR}), $\xi_{Z/\gamma}^{NR}=1.075(1\pm0.34)$. 
Putting the above discussion together, we write
\beqn
{\rm Re}\,\delta_{\gamma Z}(\nu=1.165\,{\rm GeV})&=&(5.85\pm0.45_{mod}\pm1.23_{iso})\%\;.
\eeqn
\indent
We conclude that the main uncertainty to the
dispersion correction comes from the isospin decomposition of the
electromagnetic data, most notably from high energy background. We showed how
the total uncertainty can be represented as an uncertainty in the four
isospin scaling parameters, $\xi_{Z/\gamma}^{NR}$,
$\xi_{Z/\gamma}^{P_{33}(1232)}$, $\xi_{Z/\gamma}^{S_{11}(1535)}$ 
and $\xi_{Z/\gamma}^{F_{37}(1950)}$. The estimates of 
uncertainties in these factors that we presented are only approximate,
and will be studied in greater detail in an upcoming work in order to ensure the
interpretability of the QWEAK experiment. 
\bibliographystyle{aipproc} 

\end{document}